\documentclass[a4paper,aps,onecolumn,nobibnotes,nofootinbib,pra,showpacs]{revtex4-2}

\usepackage{amssymb,amsmath,amsthm,mathtools}
\usepackage{bbm,physics,braket,mathrsfs,bbding}

\DeclareMathOperator*{\Motimes}{\text{\raisebox{0.25ex}{\scalebox{0.8}{$\bigotimes$}}}}

\begin{document}
	
	\title{Axiomatic approach to measures of total correlations}
	
	\author{Gabriel L. Moraes}
	\author{Renato M. Angelo}
	\author{Ana C. S. Costa}
	
	\affiliation{$^{1}$Department of Physics, Federal University of Paran\'a, P.O. Box 19044, 81531-980, Curitiba, Paran\'a, Brazil.}
	
	\date{\today}

\begin{abstract}
Correlations play a pivotal role in various fields of science, particularly in quantum mechanics, yet their proper quantification remains a subject of debate. In this work, we aim to discuss the challenge of defining a reliable measure of total correlations. We first outline essential properties that an effective correlation measure should satisfy and review existing measures, including quantum mutual information, the $p$-norm of the correlation matrix, and the recently defined quantum Pearson correlation coefficient. Additionally, we introduce new measures based on Rényi and Tsallis relative entropies, as well as the Kullback-Leibler divergence. Our analysis reveals that while quantum mutual information, the $p$-norm, and the Pearson measure exhibit equivalence for two-qubit systems, they all suffer from an ordering problem. Despite criticisms regarding its reliability, we argue that QMI remains a valid measure of total correlations.
\end{abstract}

\maketitle

\section{Introduction}
\label{Introduction}

Whenever a bipartite quantum system is described by a product state $\rho_{12}=\rho_1\otimes\rho_2$, the system is said to be uncorrelated. Physically, this means that the probability distributions associated with part $1$ in no way depend on those of part $2$. When such full factorization does not apply, the system is said to possess a feature called {\it correlation}. With applications in several fields~\cite{Fanchini2017}, this feature is essential in quantum foundations \cite{Hu2018,Wilde2013,Jaeger2007,Nielsen2000}, quantum information science \cite{Xi2012,Nguyen2018,Khalid2020}, and quantum thermodynamics~\cite{Goold2016,Perarnau2015,Micadei2019,Andolina2019,Alipour2016,Barrios2017}. It is well established by now that correlations can be distinguished as classical and quantum~\cite{Vedral2020}, with the latter implying advantages in certain physical tasks \cite{Modi2011,Branciard2012,He2015,Xiang2017,Goettems2021}.

One way to quantify the total correlations encoded in a quantum state is through quantum mutual information (QMI). As discussed in Ref.~\cite{Li2007}, the expression ``quantum mutual information'' dates back to the work of Cerf and Adami~\cite{Adami1997}. In contrast, its mathematical formulation, as it is known nowadays, recalls the quantity named ``coupling information'' in the work of Stratonovich~\cite{Stratonovich1966}. One of its first uses was done by Zurek while studying information in quantum measurements~\cite{Zurek1983}. Furthermore, it is also called the ``correlation index'' by Barnett and Phoenix~\cite{Barnett1989,Barnett1991}. In the context of multipartite states, it has recently been shown that at least $n-1$ families of quantum mutual information exist for a system of $n\geq 2$ parties \cite{Kumar2024}. Furthermore, any positive linear combination of these families will generate another valid family. 

Although QMI is widely used as a measure of total correlations, other measures can be found in the literature, such as the norm of the correlation matrix~\cite{Alipour2020} and the proposal based on the Pearson coefficient~\cite{Tserkis2023}. However, these measures might not always agree on the strength of the correlation, bringing up a behavior among them known as the {\sl ordering problem}. Regarding other types of correlations, such as entanglement, known measures may not consistently order states in terms of their level of correlation~\cite{Virmani2000}. In other words, different measures may not agree which one of two given quantum states is {\sl more} correlated. Because QMI also exhibits this characteristic when compared with the norm of the correlation matrix, the authors of Ref.~\cite{Alipour2020} suggest that QMI may not be a reliable measure of total correlations. Of course, one might wonder why the ordering problem is not attributed to the norm of the correlation matrix instead.

This ambiguity persists because the concept of total correlations lacks a clear characterization in terms of a minimal set of axioms that a measure should satisfy. The aim of this article is to address the question: how to define a {\it bona fide} measure of total correlations? To accomplish this task, we introduce in Sec.~\ref{ii} a list of properties we deem necessary for a proper measure. In Sec.~\ref{iii}, we present some proposals of correlation measures and, in Sec.~\ref{iv}, we analyze them for general two-qubit systems. Sec.~\ref{v} is reserved for discussions and future perspectives. As a final product, we reaffirm, based on our axiomatic approach, that QMI remains the most faithful measure of total correlations so far.

\section{Axioms}
\label{ii}
It is an open question what properties a measure of total correlation should have, although some works have already hinted at some of them~\cite{Henderson2001,Zhou2006,Kumar2024}. In what follows, using reasonability only, we postulate some properties we deem to be non-negotiable for faithful quantifier of total correlations. For future convenience, we introduce the factorizing map $\Omega$, which transforms any $N$-partite quantum state $\rho$ acting on the Hilbert space $\mathcal{H}=\mathcal{H}_1\otimes\mathcal{H}_2\otimes\cdots\otimes\mathcal{H}_N$ into its fully uncorrelated counterpart $\rho_1\otimes\rho_2\otimes\cdots\otimes\rho_N$. Symbolically,
\begin{equation}\label{Omega}
\Omega(\rho):=\bigotimes_{i=1}^{N} \Tr_{\langle i \rangle} \rho = \bigotimes_{i=1}^{N} \rho_i\equiv \Omega_\rho,
\end{equation}
where the operation $\Tr_{\langle i \rangle}$ means that every part except the $i$-th one should be traced out. The properties that ought to be satisfied by every measure $\mathcal{C}(\rho)$ of the total correlations encoded in a specific multi-partition of the full Hilbert space are as follows.

(i) {\it Non-negativity}: $\mathcal{C}(\rho) \geq 0$, with equality holding if and only if $\rho=\Omega_\rho$.

(ii) {\it Local unitary invariance}: $\mathcal{C}\big(U_\text{\tiny LOC}\,\rho\,U_\text{\tiny LOC}^\dag\big)=\mathcal{C}(\rho)$, where $U_\text{\tiny LOC}=\bigotimes_{i=1}^N U_i$ and $U_i$ are unitary transformations acting on the single Hilbert space $\mathcal{H}_i$.

(iii) {\it Monotonicity under local completely positive trace preserving (CPTP)} maps: $\mathcal{C}(\eta)\leq\mathcal{C}(\rho)$, where $\eta=\varepsilon_1\circ \varepsilon_2\circ\cdots\circ\varepsilon_N (\rho)$ and $\varepsilon_i$ is a CPTP map acting on part $i$ only. In particular, $\mathcal{C}(\Tr_i\rho)\leq \mathcal{C}(\rho)$.

(iv) {\it Continuity}: $\mathcal{C}(\rho)-\mathcal{C}(\rho_\epsilon)\geq\epsilon\,  \mathcal{C}(\rho)$ for $\rho_\epsilon\coloneqq (1-\epsilon)\,\rho+\epsilon\,\Omega_\rho$ and $ 0\leq \epsilon\leq 1$. Also, $\mathcal{C}(\rho_{1-\varepsilon})\leq \varepsilon\,\mathcal{C}(\rho)$, where $0\leq \varepsilon\leq 1$.

(v) {\it Additivity}: $\mathcal{C}(\rho_\mathcal{X}\otimes\rho_\mathcal{Y})=\mathcal{C}(\rho_\mathcal{X})+\mathcal{C}(\rho_\mathcal{Y})$, with $\rho_\mathcal{X}$ and $\rho_\mathcal{Y}$ multipartite states corresponding to the composite parts $\mathcal{X}=\{1,2,\dots,k\}$ and $\mathcal{Y}=\{k+1,k+2,\dots,N\}$, respectively.

Condition (i) implies a positive semidefinite scale for the correlations measure, with zero representing the lack of correlations. Conditions (ii) and (iii) are often required for correlation measures~\cite{Vedral2003,Li2007}. The former incorporates the fact that correlations are global properties that should not change upon unitary operations. The latter implements the concept that local CPTP maps do not increase correlations (in particular, discarding a part removes correlations). Condition (iv) requires that if one slightly degrades a correlated state $\rho$, particularly towards reaching its uncorrelated counterpart $\Omega_\rho$, then correlations will be destroyed, but only to a small extent. Similarly, if one slightly perturbs an uncorrelated state $\Omega_\rho$, particularly towards reaching its correlated counterpart $\rho$, then correlations will be generated, but only to a small extent. Since $\mathcal{C}(\Omega_\rho)=0$, it follows that condition (iv) also implies the convexity of $\mathcal{C}$ with respect to $\rho_\epsilon$. This is not to say, though, that $\mathcal{C}$ must be convex in general, for it is clear that $(1-\epsilon)\rho_1\otimes\rho_2+\epsilon \sigma_1\otimes\sigma_2\neq \omega_1\otimes\omega_2$, meaning that mixing can increase correlations. Condition (v) is justified as follows. Consider that the multipart $\mathcal{X}=\{1,2\}\equiv 12$ of a system has no physical connection whatsoever with multipart $\mathcal{Y}=\{3,4\}\equiv 34$, i.e., $\mathcal{X}$ and $\mathcal{Y}$ are actually statistically independent systems. It is natural that a quantifier detects only the remaining internal correlations within the multiparts. In particular, it must hold that $\mathcal{C}(\rho_{12} \otimes \rho_{3}) = \mathcal{C}(\rho_{12})$. 

Some remarks are now opportune. First, the five axioms proposed above are not intended to exhaust all the reasonable conditions that must be satisfied by a faithful measure of total correlations. We expect, however, that they form a set that is restrictive enough to select among candidate measures. Second, it is not our concern here to describe how a multipart $\mathcal{X}=\{1,2,\cdots,k\}$ is correlated with another multipart $\mathcal{Y}=\{k+1,k+2,\cdots,N\}$, although this is a valid problem. To accomplish this task, one could, for instance, redefine the factorizing map to yield $\Omega(\rho)=\rho_\mathcal{X}\otimes\rho_\mathcal{Y}$. Third, additivity condition is relevant only in scenarios involving different multiparts. In the simplest bipartite case ($N=2$), it makes no sense. Note that properties (i)–(iii) are similar to the axioms proposed by Henderson and Vedral in~\cite{Henderson2001}, while the axioms (iv)–(v) represent new additions. As already mentioned above, we understand that these two conditions contribute to the construction of reliable measures: condition (iv) ensures smooth functions, while condition (v) guarantees that uncorrelated states do not affect the degree of total correlations for a specific multi-partition of the total system.

\section{Total Correlation Measures}
\label{iii}
Quantifying correlations is an important and equally difficult task, as we have learned from the entanglement literature~\cite{Peres1996,Horodecki1996,Schumacher2023}. In principle, one might propose infinitely many functionals that are well-behaved mathematically but fail to satisfy a given set of reasonable axioms, as the one proposed above. The aim of this section is to make this point and, ultimately, identify faithful measures of total correlations.

\subsection{Quantum mutual information} 

The first candidate measure can be none other than the most traditional one, namely, QMI. Considering an $N$-partite state $\rho$ and its uncorrelated counterpart defined in Eq.~\eqref{Omega}, QMI can be expressed as
\begin{equation} \label{QMI}
I(\rho)=D(\rho||\Omega_\rho)
\end{equation}
where $D(\varrho||\sigma):=\Tr [\varrho(\log \varrho-\log\sigma)]$ is the Umegaki divergence between the quantum states $\varrho$ and $\sigma$~\cite{Umegaki1962}. Also known in the literature as von Neumann relative entropy, $D$ is the quantum extension of the Kullback-Leibler divergence~\cite{Cover2006}. The Umegaki divergence satisfies several interesting properties, such as positive definiteness, being null if only if its entries are equal, invariance under the action of local unitaries, monotonicity under CPTP maps (also known as contractivity), joint convexity, and additivity~\cite{Umegaki1962}. For the purposes of the analyses conducted in the present work, we refer the reader to the useful summary of properties reported in Table I of Ref.~\cite{Orthey2022}. 

Defined in terms of the Umegaki divergence, $I(\rho)$ directly satisfies all the axioms on our list, which can be demonstrated straightforwardly. To be sure, we remark that condition $\mathcal{C}(\Tr_i\rho)\leq \mathcal{C}(\rho)$ is proven by recognizing that partial tracing is a CPTP map, while condition (iv) derives from joint convexity. For practical purposes, it is interesting to note that QMI can be written in terms of the von Neumann entropy, $S(\rho) =-\Tr(\rho\log\rho)$, as
\begin{equation}\label{QMI-S}
I(\rho) = \sum_{i=1}^N S(\rho_i) - S(\rho).
\end{equation}
%

\subsection{Rényi divergence}

Our second candidate measure of total correlations is
\begin{equation}\label{RenyiD}
	I_\alpha(\rho) = D_\alpha(\rho||\Omega_\rho),
\end{equation}
where $D_\alpha(\varrho||\sigma)\coloneqq \frac{1}{1-\alpha} \log\big[\Tr \big(\varrho^\alpha\sigma^{1-\alpha}\big)\big]$ is the Rényi divergence between the quantum states $\varrho$ and $\sigma$, for $\alpha \in (0,1)\cup(1,+\infty)$~\cite{Petz1986}. The one-parameter measure \eqref{RenyiD} is a clear generalization of \eqref{QMI}, since $D_{\alpha\to1}(\varrho||\sigma)=D(\varrho||\sigma)$. Some variants of the Rényi divergence are known (see Ref.~\cite{Orthey2022} and references therein) that can also be candidate measures. However, they will not be analyzed here because our intention is to be illustrative only. All properties we mentioned that are satisfied by the Umegaki divergence remain intact for all values of $\alpha$, except monotonicity under the action of CPTP maps, which is valid only for $\alpha\in(0,1)\cup(1,2]$~\cite{Mosonyi2011}. It then follows that, as long as one restricts $\alpha$ to the aforementioned domains, $I_\alpha(\rho)$ will also be a faithful measure of total correlations.

\subsection{Tsallis relative entropy}

Another generalization of the measure~\eqref{QMI} is
\begin{equation}\label{TsallisD}
I_q(\rho)=D_q(\rho||\Omega_\rho), 
\end{equation}
where $D_q(\varrho||\sigma)\coloneqq \Tr\big[\varrho^q(\ln_q\varrho-\ln_q\sigma)\big]$ is the Tsallis relative entropy of the quantum states $\varrho$ and $\sigma$~\cite{Abe2003,Rastegin2016}, with $q \in (0,1)\cup(1,+\infty)$ and $\ln_q(x):=(x^{1-q}-1)/(1-q)$. It is well known that $D_{q\to 1}(\varrho||\sigma)=D(\varrho||\sigma)$. Once again, the logic behind the measure is to quantify how distinguishable $\rho$ is from its uncorrelated counterpart $\Omega_\rho$. Although the Tsallis relative entropy is regarded as an extension of von Neumann relative entropy, it does not preserve all the properties listed in Sec.~\ref{ii}. From the results reported in Refs.~\cite{Rastegin2016,Hiai2011}, one can validate properties (i) and (ii) for the entire domain of $q$, and properties (iii) and (iv) for  $q\in(0,1)\cup(1,2]$, but additivity is violated. Recall, however, that this does not constitute a problem for $N=2$, meaning that $I_q(\rho)$ is a faithful measure for bipartite states.

\subsection{Norm of correlation matrix}

So far, we have considered only entropy-based measures, comparing $\rho$ with $\Omega_\rho$. We now analyze the adequacy of a norm-based measure. Specifically, drawing inspiration from Ref.~\cite{Alipour2020}, we consider the ``geometric quantifier''
\begin{equation}\label{SchattenD}
\mathscr{G}_p(\rho)=||\rho-\Omega_\rho||_p,
\end{equation}
where $||O||_p\coloneqq \big[\Tr \big(O^\dag O\big)^{p/2}\big]^{1/p}$ is the Schatten $p$-norm of an operator $O$, for any real number $p\geq 1$, and $\rho-\Omega_\rho$ is the correlation matrix. From the semipositive definiteness of any norm, it follows that condition (i) is promptly satisfied. Property (ii) is guaranteed by the well-known fact that the Schatten $p$-norm is unitary invariant~\cite{Bhatia2013}. Condition (iii) is satisfied, but only under a special condition. In Ref.~\cite{Alipour2020}, the authors used the 2-norm to define the correlation matrix. However, for this specific value, monotonicity is satisfied only for unital CPTP maps~\cite{Bhatia2013}, which is ensured by the Pinching inequality. Here, however, we are interested in monotonicity being satisfied for any CPTP map, and the contractivity of the norms holds only for $p=1$. Because all norms are convex (by the triangle inequality), $\mathscr{G}_p(\rho)$ automatically satisfies condition (iv). More specifically, since $\Omega_{\rho_\epsilon}=\Omega_\rho$, one has $\mathscr{G}_p(\rho_\epsilon)=(1-\epsilon)\mathscr{G}_p(\rho)$. To show that additivity is not respected by $\mathscr{G}_p$, let us consider, for simplicity, the state $\rho_{12}\otimes\rho_3$. Plugging this state and its uncorrelated counterpart $\rho_1\otimes\rho_2\otimes\rho_3$ into Eq.~\eqref{SchattenD} returns $\mathscr{G}_p(\rho_{12}\otimes\rho_3)=\mathscr{G}_p(\rho_{12}) \,||\rho_3||_p$. It is then clear that the measure under scrutiny has an important flaw: the correlations encoded in $\rho_{12} \otimes \rho_3$ are given not only by the correlations in $\rho_{12}$, but also by the norm of the completely irrelevant state $\rho_3$. Therefore, although the measure proposed in Ref.~\cite{Alipour2020} is fine for bipartite states, it does not pass the axiomatic test we introduced in Sec.~\ref{ii} for $N>2$. One might suspect that normalizing the formula \eqref{SchattenD} with $||\Omega_\rho||_p$ would fix the problem, but this is not true. Doing so would instead affect the other properties.

\subsection{Pearson correlation coefficient}

For the classical case, it has been shown that the Pearson correlation coefficient (PCC)---a measure based on the first and second-moment statistics of distributions---can satisfactorily capture the correlation between two random variables~\cite{Pearson1895,Renyi1970}. Recently, an extension of the Pearson coefficient was introduced in the quantum domain to formulate a measure of total correlations for bipartite states~\cite{Tserkis2023}. Here, we introduce a generalization of the concept for the $N$-partite case.

Let $W_i$ be a discrete-spectrum observable acting on a Hilbert space $\mathcal{H}_i$. Using the notation $\mathbf{W}=\{W_1,W_2,\cdots,W_N\}$, we consider the following covariance function:
\begin{align}\label{CW}
\mathscr{C}_{\rho}^\mathbf{W}\coloneqq & \Tr\Big[\Motimes_{i=1}^N W_i\,\,\big(\rho-\Omega_\rho\big) \Big] \\
=& \langle W_1\otimes W_2\otimes \cdots \otimes W_N \rangle_\rho-\langle W_1\rangle_{\rho_1}\langle W_2\rangle_{\rho_2}\cdots \langle W_N\rangle_{\rho_N},\nonumber
\end{align}
Let $\Delta W_{i,\rho}=\big(\langle W_i^2\rangle_\rho-\langle W_i\rangle_\rho^2\big)^{1/2}$ be the uncertainty of $W_i$, where $\langle W_i^n\rangle_\rho=\Tr\big(W_i^n\rho\big)$ for $n\in\mathbbm{N}_{>0}$. The $N$-partite generalization of PCC then reads
\begin{equation}
\mathscr{R}^\mathbf{W}(\rho):=\frac{\mathscr{C}_\rho^{\mathbf{W}}}{\Motimes_i\Delta W_{i,\rho}}.
\end{equation}
The normalization guarantees that $\mathscr{R}^\mathbf{W}(\rho)$ is dimensionless. In order to capture all existing correlations in the quantum state with the smallest number of measurements, the authors in Ref.~\cite{Tserkis2023} considered the maximum set of complementary observables in the Hilbert space of each subsystem. To implement this procedure here, we have to replace each operator $W_i$ in the construction above with $W_i^k$, where $k\in\{1,2,\cdots,v\}$ and $v$ is the cardinality of the maximum set of {\it pairwise complementary observables}\footnote{Two maximally incompatible observables $X$ and $Y$ are called complementary, denoted as $X\natural Y$ , and have mutually unbiased bases (MUB), i.e., $|\langle x_i|y_j\rangle|^2=1/d$ $\forall\,i,j\in\{0,1,\cdots,d-1\}$.} $W_i^k \natural W_i^{k'}$ acting on $\mathcal{H}_i$. Accordingly, in what follows, the set $\mathbf{W}$ is updated to $\mathbf{W}^k=\{W_1^k,W_2^k,\cdots,W_N^k\}$. Thus, the total correlations of an $N$-partite state $\rho$ can be calculated using the PCC measure as
\begin{equation}\label{R}
\mathbb{R}(\rho):=\max_{\mathbf{W}^k}\sum^v_{k=1}\big|\mathscr{R}^{\mathbf{W}^k}(\rho)\big|,  
\end{equation}
where the maximization is taken over all possible sets $\mathbf{W}^k$.

The measure \eqref{R} satisfies axioms (i) and (ii) by construction: the former via the modulus, and the latter via the maximization. It also satisfies monotonicity under CPTP maps~\cite{Beigi2013,Beigi2023}. Through direct calculations one proves that $\mathbb{R}(\rho_\epsilon)=(1-\epsilon)\,\mathbb{R}(\rho)$, thus showing that axiom (iv) is satisfied. Finally, it is not difficult to show that the additivity axiom is violated for $N>2$.

\subsection{Kullback-Leibler divergence}

As in the previous section, here we investigate a measure starting with correlations associated with local measurements. Let $Z_i=\sum_z z_i Z_z^i$ be a discrete-spectrum observable acting on a Hilbert space $\mathcal{H}_i$, where $Z_z^i$ are projection operators satisfying $Z_z^iZ_{z'}^i=\delta_{zz'}Z_z^i$ and $z_i$ are eigenvalues of $Z_i$. The key idea here is to quantify how distinguishable the joint probability distribution
\begin{equation}
p(\mathbf{z}|\rho)\equiv p(z_1,z_2,\cdots,z_n|\rho)=\Tr\left(\Motimes_{i=1}^NZ_z^i\,\,\rho\right)
\end{equation}
is from $p(\mathbf{z}|\Omega_\rho)$. Once again we can resort to an entropy-based tool [here, the Kullback-Leibler (KL) divergence and all its known properties] to measure the total correlations:
\begin{equation}\label{D_KL}
D_\text{\tiny KL}(\rho)=\max_{\{Z_z^i\}}\sum_\mathbf{z}p(\mathbf{z}|\rho)\log\left[\frac{p(\mathbf{z}|\rho)}{p(\mathbf{z}|\Omega_\rho)}\right].
\end{equation}

Axiom (i) is satisfied because KL divergence is positive semidefinite, being null iff its entries are equal. To verify the validity of axiom (ii), we first note that
\begin{equation}
\Tr\left(\Motimes_i Z_z^i\,U_\text{\tiny LOC}\,\rho\,U_\text{\tiny LOC}^\dag \right)=\Tr\left(\Motimes_i \tilde{Z}_z^i\,\rho \right),
\end{equation}
with $\tilde{Z}_z^i=U_i^\dag Z_z^iU_i$. Because the maximization in Eq.~\eqref{D_KL} is supposed to run across all projection operators, the $\tilde{Z}_z^i$ are also considered. It follows that $D_\text{\tiny KL}\big(U_\text{\tiny LOC}\,\rho\,U_\text{\tiny LOC}^\dag\big)=D_\text{\tiny KL}(\rho)$, as desired. Axiom (iii) is satisfied given that the KL divergence is monotonic under the action of CPTP maps~\cite{Gour2021}. Axiom (iv) directly follows from the convexity of the KL divergence and straightforward calculations yields $D_\text{\tiny KL}(\rho_\mathcal{X}\otimes\rho_\mathcal{Y})=D_\text{\tiny KL}(\rho_\mathcal{X})+D_\text{\tiny KL}(\rho_\mathcal{Y})$, thus validating axiom (v).

Note that the idea of investigating correlations through the analysis of the effect of local measurements is widely used in the literature, for example, in the definition of symmetric quantum discord~\cite{Maziero2011}.

\section{Analysis of the total correlation measures for two-qubit systems}
\label{iv}

\begin{figure}
\centering
\includegraphics[scale=0.25]{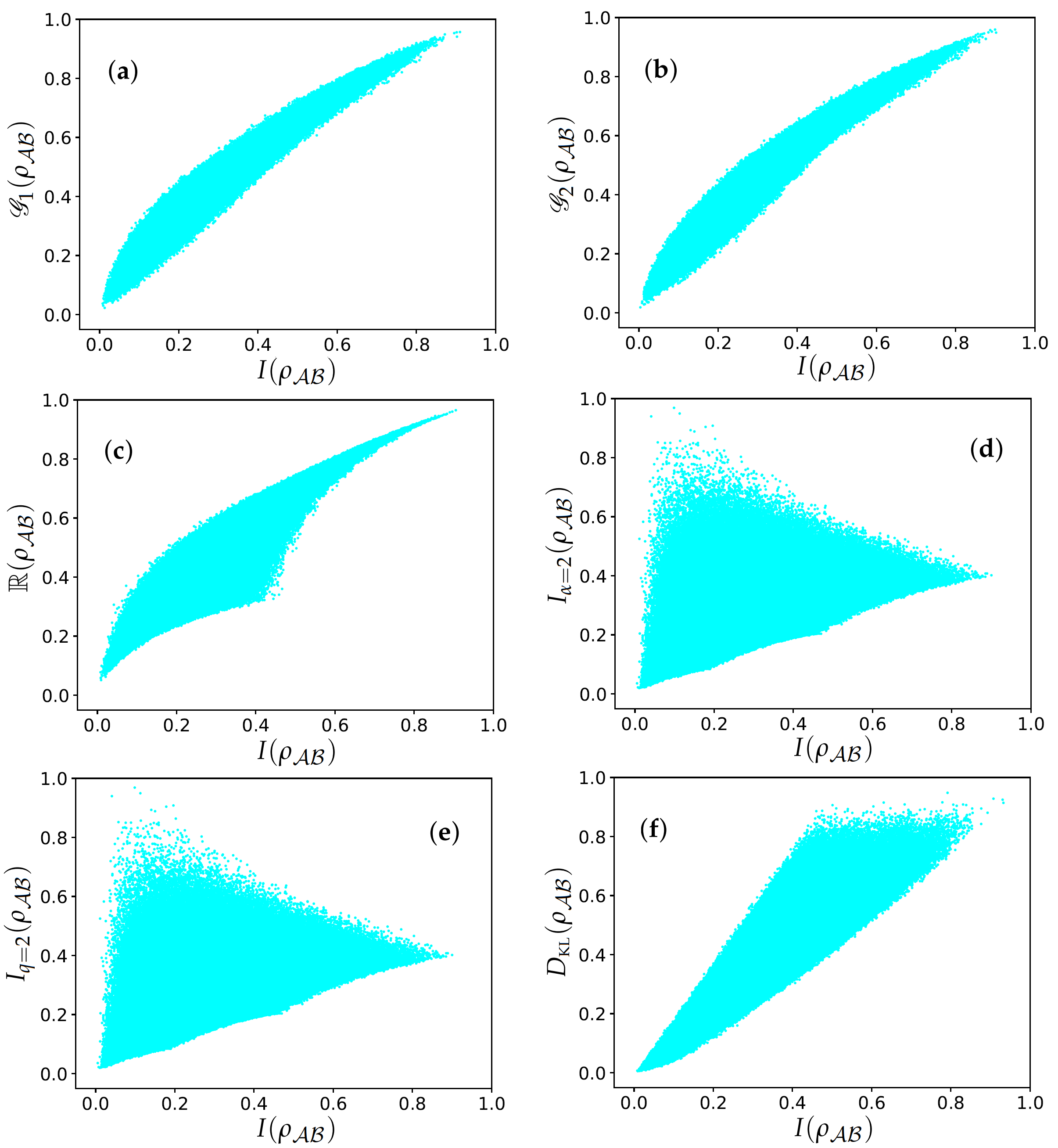}
\caption{
Diagrams of (${\bf a}$) $\mathscr{G}_1(\rho_{\mathcal{AB}})$, (${\bf b}$) $\mathscr{G}_2(\rho_{\mathcal{AB}})$, (${\bf c}$) $\mathbb{R}(\rho_{\mathcal{AB}})$, (${\bf d}$) $I_{\alpha=2}(\rho_{\mathcal{AB}})$, (${\bf e}$) $I_{q=2}(\rho_{\mathcal{AB}})$, and (${\bf f}$) $D_\text{\tiny KL}(\rho_{\mathcal{AB}})$ with $I(\rho_{\mathcal{AB}})$. Every point represents one of the $10^6$ random quantum states generated for each one of the panels. For the plots, we chose $p=1$, $p=2$, $\alpha=1$, and $q=2$ for simplicity. All measures are normalized by their maximal value restricting them to the interval [0,1].} 
\label{fig1}
\end{figure}

In this section, we aim to analyze the presented measures of total correlations for systems of two qubits. To do so, we compared all of them with QMI. The latter is chosen as the reference given its importance in the literature and because it satisfies all required properties of a well-behaved measure. For comparison, we generated $10^6$ random quantum states and confronted the measures as shown in Fig.~\ref{fig1}. In these comparisons, we expect equivalence between the measures; that is, for the states where QMI is at a maximum (minimum), the corresponding proposal in the comparison should also be at a maximum (minimum).

In Fig.~\ref{fig1}, one can see that $\mathscr{G}_p(\rho_{\mathcal{AB}})$ (where we chose the specific norms 1 and 2), and $\mathbb{R}(\rho_{\mathcal{AB}})$ are equivalent to $I(\rho)$. This indicates that they can be considered good measures of total correlations when considering two-qubit systems. However, $I_{\alpha=2}(\rho_{\mathcal{AB}})$, $I_{q=2}(\rho_{\mathcal{AB}})$, and $D_\text{\tiny KL}(\rho_\mathcal{AB})$ agree about the minimum while diverging about the quantum state that maximizes the measure. 
Although the plots display only specific values of $p$, $\alpha$, and $q$, the same behavior was observed for other values of these parameters, provided that the relevant properties are satisfied. Specifically, we restrict $\alpha$ and $q$ to the range $(0,2]\setminus \{1\}$. Finally, for $\mathscr{G}_p$, we tested the value $p=1$, for which the condition (iii) is satisfied, and $p=2$, as originally investigated in~\cite{Alipour2020}.

Regarding the properties that have been enunciated as necessary for a measure of total correlations, except for Tsallis relative entropy-based measures~\eqref{TsallisD} and the $p$-norm of the correlation matrix~\eqref{SchattenD}, which fail to satisfy condition (iii), the other proposed measures satisfy all of them. This result somehow diverges from the one presented in~\cite{Alipour2020}, where the authors argue in favor of the norm of the correlation matrix as a reliable measure of total correlations. It is important to notice that their criticism is solely based on the problem of ordering between measures, something that occurs for several measures of quantum resources. For completeness, Table~\ref{table} shows the reader the properties satisfied by each measure.
\begin{table}
\begin{center}
\scalebox{0.9}{ 
\begin{tabular}{c c c c c c c}
\hline
 & $I$ & $I_\alpha$ & $I_q$& $\mathscr{G}_p$ & $\mathbb{R}$ & $D_\text{\tiny KL}$\\
\hline
Non-negativity & \CheckmarkBold & \CheckmarkBold & \CheckmarkBold & \CheckmarkBold & \CheckmarkBold & \CheckmarkBold \\
Local unitary invariance & \CheckmarkBold & \CheckmarkBold & \CheckmarkBold &  \CheckmarkBold &  \CheckmarkBold & \CheckmarkBold \\
Monotonicity & \CheckmarkBold & $\alpha\in (0,2]\setminus \{1\}$ &$q\in (0,2]\setminus \{1\}$ &  $p=1$ &  \CheckmarkBold & \CheckmarkBold \\
Continuity & \CheckmarkBold & \CheckmarkBold & \CheckmarkBold & \CheckmarkBold &  \CheckmarkBold & \CheckmarkBold \\
Additivity & \CheckmarkBold & \CheckmarkBold & \XSolidBrush & \XSolidBrush  &  $N=2$ & \CheckmarkBold \\
\hline
\end{tabular}}
\caption{\label{table} Summary of properties satisfied by QMI ($I$), Rényi divergence based measure ($I_\alpha$), Tsallis relative entropies based measure ($I_q$), $p$-norm of the correlation matrix ($\mathscr{G}_p$), PCC measure ($\mathbb{R}$), and Kullback-Leibler divergence ($D_\text{\tiny KL}$).}
\end{center}
\end{table}

\section{Final remarks}
\label{v}

The importance of correlations in science, and particularly in quantum mechanics, need not be emphasized. Still, the proper quantification of these objects is repeatedly questioned, revealing that it is essential to establish axioms for correlation measures, especially for total correlations. The objective of this work is to take some steps toward deepening the issue.

This work provides a brief overview of the properties we consider essential for an effective correlation measure. Following this discussion, we present several total correlation measures from the literature, including the QMI [Eq.~\eqref{QMI}], the $p$-norm of the correlation matrix [Eq.~\eqref{SchattenD}], and the recently defined PCC measure [Eq.~\eqref{R}]. Additionally, we introduce new measures based on the Rényi [Eq.~\eqref{RenyiD}] and Tsallis [Eq.~\eqref{TsallisD}] relative entropies, as well as the Kullback-Leibler divergence [Eq.~\eqref{D_KL}].

While analyzing the properties satisfied by each measure, we found that additivity is not fulfilled by the $p$-norm of the correlation matrix and the Tsallis-based measure. This aspect raises the question of how essential this property is for a correlation measure. However, if one chooses to disregard this property, the measure's sensitivity to uncorrelated systems may be affected, although its equivalence with the QMI remains unchanged. In this context, among the measures considered, we demonstrated that for systems of two qubits, only the QMI, the $p$-norm of the correlation matrix, and the PCC measure are equivalent to each other. Nevertheless, they all exhibit an ordering problem. All in all, our work suggests that, despite some claims to the contrary~\cite{Alipour2020}, QMI remains an arguably reliable measure of total correlations.

For future work, it would be interesting to investigate monogamy relations for the introduced measures, showing how total correlations spread in this kind of system for each of the considered measures. 

\section{Acknowledgments}
G.L.M. acknowledges support by the CNPq/Brazil. R.M.A. thanks the financial support from the National Institute for Science and Technology of Quantum Information (CNPq, INCT-IQ 465469/2014-0). R.M.A and A.C.S.C thank the Brazilian funding agency CNPq under Grants No. 305957/2023-6 and 308730/2023-2, respectively.

\bibliography{references}

\begin{thebibliography}{51}%
\makeatletter
\providecommand \@ifxundefined [1]{%
 \@ifx{#1\undefined}
}%
\providecommand \@ifnum [1]{%
 \ifnum #1\expandafter \@firstoftwo
 \else \expandafter \@secondoftwo
 \fi
}%
\providecommand \@ifx [1]{%
 \ifx #1\expandafter \@firstoftwo
 \else \expandafter \@secondoftwo
 \fi
}%
\providecommand \natexlab [1]{#1}%
\providecommand \enquote  [1]{``#1''}%
\providecommand \bibnamefont  [1]{#1}%
\providecommand \bibfnamefont [1]{#1}%
\providecommand \citenamefont [1]{#1}%
\providecommand \href@noop [0]{\@secondoftwo}%
\providecommand \href [0]{\begingroup \@sanitize@url \@href}%
\providecommand \@href[1]{\@@startlink{#1}\@@href}%
\providecommand \@@href[1]{\endgroup#1\@@endlink}%
\providecommand \@sanitize@url [0]{\catcode `\\12\catcode `\$12\catcode
  `\&12\catcode `\#12\catcode `\^12\catcode `\_12\catcode `\%12\relax}%
\providecommand \@@startlink[1]{}%
\providecommand \@@endlink[0]{}%
\providecommand \url  [0]{\begingroup\@sanitize@url \@url }%
\providecommand \@url [1]{\endgroup\@href {#1}{\urlprefix }}%
\providecommand \urlprefix  [0]{URL }%
\providecommand \Eprint [0]{\href }%
\providecommand \doibase [0]{https://doi.org/}%
\providecommand \selectlanguage [0]{\@gobble}%
\providecommand \bibinfo  [0]{\@secondoftwo}%
\providecommand \bibfield  [0]{\@secondoftwo}%
\providecommand \translation [1]{[#1]}%
\providecommand \BibitemOpen [0]{}%
\providecommand \bibitemStop [0]{}%
\providecommand \bibitemNoStop [0]{.\EOS\space}%
\providecommand \EOS [0]{\spacefactor3000\relax}%
\providecommand \BibitemShut  [1]{\csname bibitem#1\endcsname}%
\let\auto@bib@innerbib\@empty
\bibitem [{\citenamefont {Fanchini}\ \emph {et~al.}(2017)\citenamefont
  {Fanchini}, \citenamefont {Pinto},\ and\ \citenamefont
  {Adesso}}]{Fanchini2017}%
  \BibitemOpen
  \bibfield  {author} {\bibinfo {author} {\bibfnamefont {F.~F.}\ \bibnamefont
  {Fanchini}}, \bibinfo {author} {\bibfnamefont {D.~O.~S.}\ \bibnamefont
  {Pinto}},\ and\ \bibinfo {author} {\bibfnamefont {G.}~\bibnamefont
  {Adesso}},\ }\href@noop {} {\emph {\bibinfo {title} {Lectures on general
  quantum correlations and their applications}}}\ (\bibinfo  {publisher}
  {Springer},\ \bibinfo {year} {2017})\BibitemShut {NoStop}%
\bibitem [{\citenamefont {Hu}\ \emph {et~al.}(2018)\citenamefont {Hu},
  \citenamefont {Hu}, \citenamefont {Wang}, \citenamefont {Peng}, \citenamefont
  {Zhang},\ and\ \citenamefont {Fan}}]{Hu2018}%
  \BibitemOpen
  \bibfield  {author} {\bibinfo {author} {\bibfnamefont {M.}~\bibnamefont
  {Hu}}, \bibinfo {author} {\bibfnamefont {X.}~\bibnamefont {Hu}}, \bibinfo
  {author} {\bibfnamefont {J.}~\bibnamefont {Wang}}, \bibinfo {author}
  {\bibfnamefont {Y.}~\bibnamefont {Peng}}, \bibinfo {author} {\bibfnamefont
  {Y.}~\bibnamefont {Zhang}},\ and\ \bibinfo {author} {\bibfnamefont
  {H.}~\bibnamefont {Fan}},\ }\bibfield  {title} {\bibinfo {title} {Quantum
  coherence and geometric quantum discord},\ }\href
  {https://doi.org/10.1016/j.physrep.2018.07.004} {\bibfield  {journal}
  {\bibinfo  {journal} {Phys. Rep.}\ }\textbf {\bibinfo {volume} {762-764}},\
  \bibinfo {pages} {1} (\bibinfo {year} {2018})}\BibitemShut {NoStop}%
\bibitem [{\citenamefont {Wilde}(2013)}]{Wilde2013}%
  \BibitemOpen
  \bibfield  {author} {\bibinfo {author} {\bibfnamefont {M.~M.}\ \bibnamefont
  {Wilde}},\ }\href@noop {} {\emph {\bibinfo {title} {Quantum information
  theory}}}\ (\bibinfo  {publisher} {Cambridge University Press},\ \bibinfo
  {year} {2013})\BibitemShut {NoStop}%
\bibitem [{\citenamefont {Jaeger}(2007)}]{Jaeger2007}%
  \BibitemOpen
  \bibfield  {author} {\bibinfo {author} {\bibfnamefont {G.}~\bibnamefont
  {Jaeger}},\ }\href@noop {} {\emph {\bibinfo {title} {Quantum information}}}\
  (\bibinfo  {publisher} {Springer New York},\ \bibinfo {year}
  {2007})\BibitemShut {NoStop}%
\bibitem [{\citenamefont {Nielsen}\ and\ \citenamefont
  {Chuang}(2000)}]{Nielsen2000}%
  \BibitemOpen
  \bibfield  {author} {\bibinfo {author} {\bibfnamefont {M.~A.}\ \bibnamefont
  {Nielsen}}\ and\ \bibinfo {author} {\bibfnamefont {I.~L.}\ \bibnamefont
  {Chuang}},\ }\href@noop {} {\emph {\bibinfo {title} {Quantum Computation and
  Quantum Information}}}\ (\bibinfo  {publisher} {Cambridge University Press},\
  \bibinfo {year} {2000})\BibitemShut {NoStop}%
\bibitem [{\citenamefont {Xi}\ \emph {et~al.}(2012)\citenamefont {Xi},
  \citenamefont {Wang},\ and\ \citenamefont {Li}}]{Xi2012}%
  \BibitemOpen
  \bibfield  {author} {\bibinfo {author} {\bibfnamefont {Z.}~\bibnamefont
  {Xi}}, \bibinfo {author} {\bibfnamefont {X.}~\bibnamefont {Wang}},\ and\
  \bibinfo {author} {\bibfnamefont {Y.}~\bibnamefont {Li}},\ }\bibfield
  {title} {\bibinfo {title} {Measurement-induced nonlocality based on the
  relative entropy},\ }\href {https://doi.org/10.1103/PhysRevA.85.042325}
  {\bibfield  {journal} {\bibinfo  {journal} {Phys. Rev. A}\ }\textbf {\bibinfo
  {volume} {85}},\ \bibinfo {pages} {042325} (\bibinfo {year}
  {2012})}\BibitemShut {NoStop}%
\bibitem [{\citenamefont {Nguyen}\ \emph {et~al.}(2018)\citenamefont {Nguyen},
  \citenamefont {Milne}, \citenamefont {Vu},\ and\ \citenamefont
  {Jevtic}}]{Nguyen2018}%
  \BibitemOpen
  \bibfield  {author} {\bibinfo {author} {\bibfnamefont {H.~C.}\ \bibnamefont
  {Nguyen}}, \bibinfo {author} {\bibfnamefont {A.}~\bibnamefont {Milne}},
  \bibinfo {author} {\bibfnamefont {T.}~\bibnamefont {Vu}},\ and\ \bibinfo
  {author} {\bibfnamefont {S.}~\bibnamefont {Jevtic}},\ }\bibfield  {title}
  {\bibinfo {title} {Quantum steering with positive operator valued measures},\
  }\href {https://doi.org/10.1088/1751-8121/aad115} {\bibfield  {journal}
  {\bibinfo  {journal} {J. Phys. A Math.}\ }\textbf {\bibinfo {volume} {51}},\
  \bibinfo {pages} {355302} (\bibinfo {year} {2018})}\BibitemShut {NoStop}%
\bibitem [{\citenamefont {Khalid}\ \emph {et~al.}(2020)\citenamefont {Khalid},
  \citenamefont {ur~Rehman},\ and\ \citenamefont {Shin}}]{Khalid2020}%
  \BibitemOpen
  \bibfield  {author} {\bibinfo {author} {\bibfnamefont {U.}~\bibnamefont
  {Khalid}}, \bibinfo {author} {\bibfnamefont {J.}~\bibnamefont {ur~Rehman}},\
  and\ \bibinfo {author} {\bibfnamefont {H.}~\bibnamefont {Shin}},\ }\bibfield
  {title} {\bibinfo {title} {Measurement-based quantum correlations for quantum
  information processing},\ }\bibfield  {journal} {\bibinfo  {journal}
  {Scientific Reports}\ }\textbf {\bibinfo {volume} {10}},\ \href
  {https://doi.org/10.1038/s41598-020-59220-y} {10.1038/s41598-020-59220-y}
  (\bibinfo {year} {2020})\BibitemShut {NoStop}%
\bibitem [{\citenamefont {Goold}\ \emph {et~al.}(2016)\citenamefont {Goold},
  \citenamefont {Huber}, \citenamefont {Riera}, \citenamefont {Rio},\ and\
  \citenamefont {Skrzypczyk}}]{Goold2016}%
  \BibitemOpen
  \bibfield  {author} {\bibinfo {author} {\bibfnamefont {J.}~\bibnamefont
  {Goold}}, \bibinfo {author} {\bibfnamefont {M.}~\bibnamefont {Huber}},
  \bibinfo {author} {\bibfnamefont {A.}~\bibnamefont {Riera}}, \bibinfo
  {author} {\bibfnamefont {L.}~\bibnamefont {Rio}},\ and\ \bibinfo {author}
  {\bibfnamefont {P.}~\bibnamefont {Skrzypczyk}},\ }\bibfield  {title}
  {\bibinfo {title} {The role of quantum information in thermodynamics — a
  topical review},\ }\href {https://doi.org/10.1088/1751-8113/49/14/143001}
  {\bibfield  {journal} {\bibinfo  {journal} {J. Phys. A Math.}\ }\textbf
  {\bibinfo {volume} {49}},\ \bibinfo {pages} {143001} (\bibinfo {year}
  {2016})}\BibitemShut {NoStop}%
\bibitem [{\citenamefont {Perarnau-Llobet}\ \emph {et~al.}(2015)\citenamefont
  {Perarnau-Llobet}, \citenamefont {Hovhannisyan}, \citenamefont {Huber},
  \citenamefont {Skrzypczyk}, \citenamefont {Brunner},\ and\ \citenamefont
  {Ac\'{\i}n}}]{Perarnau2015}%
  \BibitemOpen
  \bibfield  {author} {\bibinfo {author} {\bibfnamefont {M.}~\bibnamefont
  {Perarnau-Llobet}}, \bibinfo {author} {\bibfnamefont {K.~V.}\ \bibnamefont
  {Hovhannisyan}}, \bibinfo {author} {\bibfnamefont {M.}~\bibnamefont {Huber}},
  \bibinfo {author} {\bibfnamefont {P.}~\bibnamefont {Skrzypczyk}}, \bibinfo
  {author} {\bibfnamefont {N.}~\bibnamefont {Brunner}},\ and\ \bibinfo {author}
  {\bibfnamefont {A.}~\bibnamefont {Ac\'{\i}n}},\ }\bibfield  {title} {\bibinfo
  {title} {Extractable work from correlations},\ }\href
  {https://doi.org/10.1103/PhysRevX.5.041011} {\bibfield  {journal} {\bibinfo
  {journal} {Phys. Rev. X}\ }\textbf {\bibinfo {volume} {5}},\ \bibinfo {pages}
  {041011} (\bibinfo {year} {2015})}\BibitemShut {NoStop}%
\bibitem [{\citenamefont {Micadei}\ \emph {et~al.}(2019)\citenamefont
  {Micadei}, \citenamefont {Peterson}, \citenamefont {Souza}, \citenamefont
  {Sarthour}, \citenamefont {Oliveira}, \citenamefont {Landi}, \citenamefont
  {Batalh{\~a}o}, \citenamefont {Serra},\ and\ \citenamefont
  {Lutz}}]{Micadei2019}%
  \BibitemOpen
  \bibfield  {author} {\bibinfo {author} {\bibfnamefont {K.}~\bibnamefont
  {Micadei}}, \bibinfo {author} {\bibfnamefont {J.~P.~S.}\ \bibnamefont
  {Peterson}}, \bibinfo {author} {\bibfnamefont {A.~M.}\ \bibnamefont {Souza}},
  \bibinfo {author} {\bibfnamefont {R.~S.}\ \bibnamefont {Sarthour}}, \bibinfo
  {author} {\bibfnamefont {I.~S.}\ \bibnamefont {Oliveira}}, \bibinfo {author}
  {\bibfnamefont {G.~T.}\ \bibnamefont {Landi}}, \bibinfo {author}
  {\bibfnamefont {T.~B.}\ \bibnamefont {Batalh{\~a}o}}, \bibinfo {author}
  {\bibfnamefont {R.~M.}\ \bibnamefont {Serra}},\ and\ \bibinfo {author}
  {\bibfnamefont {E.}~\bibnamefont {Lutz}},\ }\bibfield  {title} {\bibinfo
  {title} {Reversing the direction of heat flow using quantum correlations},\
  }\href {https://doi.org/10.1038/s41467-019-10333-7} {\bibfield  {journal}
  {\bibinfo  {journal} {Nat. commun.}\ }\textbf {\bibinfo {volume} {10}},\
  \bibinfo {pages} {2456} (\bibinfo {year} {2019})}\BibitemShut {NoStop}%
\bibitem [{\citenamefont {Andolina}\ \emph {et~al.}(2019)\citenamefont
  {Andolina}, \citenamefont {Keck}, \citenamefont {Mari}, \citenamefont
  {Campisi}, \citenamefont {Giovannetti},\ and\ \citenamefont
  {Polini}}]{Andolina2019}%
  \BibitemOpen
  \bibfield  {author} {\bibinfo {author} {\bibfnamefont {G.~M.}\ \bibnamefont
  {Andolina}}, \bibinfo {author} {\bibfnamefont {M.}~\bibnamefont {Keck}},
  \bibinfo {author} {\bibfnamefont {A.}~\bibnamefont {Mari}}, \bibinfo {author}
  {\bibfnamefont {M.}~\bibnamefont {Campisi}}, \bibinfo {author} {\bibfnamefont
  {V.}~\bibnamefont {Giovannetti}},\ and\ \bibinfo {author} {\bibfnamefont
  {M.}~\bibnamefont {Polini}},\ }\bibfield  {title} {\bibinfo {title}
  {Extractable work, the role of correlations, and asymptotic freedom in
  quantum batteries},\ }\href {https://doi.org/10.1103/PhysRevLett.122.047702}
  {\bibfield  {journal} {\bibinfo  {journal} {Phys. Rev. Lett.}\ }\textbf
  {\bibinfo {volume} {122}},\ \bibinfo {pages} {047702} (\bibinfo {year}
  {2019})}\BibitemShut {NoStop}%
\bibitem [{\citenamefont {Alipour}\ \emph {et~al.}(2016)\citenamefont
  {Alipour}, \citenamefont {Benatti}, \citenamefont {Bakhshinezhad},
  \citenamefont {Afsary}, \citenamefont {Marcantoni},\ and\ \citenamefont
  {Rezakhani}}]{Alipour2016}%
  \BibitemOpen
  \bibfield  {author} {\bibinfo {author} {\bibfnamefont {S.}~\bibnamefont
  {Alipour}}, \bibinfo {author} {\bibfnamefont {F.}~\bibnamefont {Benatti}},
  \bibinfo {author} {\bibfnamefont {F.}~\bibnamefont {Bakhshinezhad}}, \bibinfo
  {author} {\bibfnamefont {M.}~\bibnamefont {Afsary}}, \bibinfo {author}
  {\bibfnamefont {S.}~\bibnamefont {Marcantoni}},\ and\ \bibinfo {author}
  {\bibfnamefont {A.~T.}\ \bibnamefont {Rezakhani}},\ }\bibfield  {title}
  {\bibinfo {title} {Extractable work, the role of correlations, and asymptotic
  freedom in quantum batteries},\ }\href {https://doi.org/10.1038/srep35568}
  {\bibfield  {journal} {\bibinfo  {journal} {Phys. Rev. Lett.}\ }\textbf
  {\bibinfo {volume} {6}},\ \bibinfo {pages} {35568} (\bibinfo {year}
  {2016})}\BibitemShut {NoStop}%
\bibitem [{\citenamefont {Barrios}\ \emph {et~al.}(2017)\citenamefont
  {Barrios}, \citenamefont {Albarr\'an-Arriagada}, \citenamefont
  {C\'ardenas-L\'opez}, \citenamefont {Romero},\ and\ \citenamefont
  {Retamal}}]{Barrios2017}%
  \BibitemOpen
  \bibfield  {author} {\bibinfo {author} {\bibfnamefont {G.~A.}\ \bibnamefont
  {Barrios}}, \bibinfo {author} {\bibfnamefont {F.}~\bibnamefont
  {Albarr\'an-Arriagada}}, \bibinfo {author} {\bibfnamefont {F.~A.}\
  \bibnamefont {C\'ardenas-L\'opez}}, \bibinfo {author} {\bibfnamefont
  {G.}~\bibnamefont {Romero}},\ and\ \bibinfo {author} {\bibfnamefont {J.~C.}\
  \bibnamefont {Retamal}},\ }\bibfield  {title} {\bibinfo {title} {Role of
  quantum correlations in light-matter quantum heat engines},\ }\href
  {https://doi.org/10.1103/PhysRevA.96.052119} {\bibfield  {journal} {\bibinfo
  {journal} {Phys. Rev. A}\ }\textbf {\bibinfo {volume} {96}},\ \bibinfo
  {pages} {052119} (\bibinfo {year} {2017})}\BibitemShut {NoStop}%
\bibitem [{\citenamefont {Vedral}(2002)}]{Vedral2020}%
  \BibitemOpen
  \bibfield  {author} {\bibinfo {author} {\bibfnamefont {V.}~\bibnamefont
  {Vedral}},\ }\bibfield  {title} {\bibinfo {title} {The role of relative
  entropy in quantum information theory},\ }\href
  {https://doi.org/10.1103/RevModPhys.74.197} {\bibfield  {journal} {\bibinfo
  {journal} {Rev. Mod. Phys.}\ }\textbf {\bibinfo {volume} {74}},\ \bibinfo
  {pages} {197} (\bibinfo {year} {2002})}\BibitemShut {NoStop}%
\bibitem [{\citenamefont {Modi}\ \emph {et~al.}(2011)\citenamefont {Modi},
  \citenamefont {Cable}, \citenamefont {Williamson},\ and\ \citenamefont
  {Vedral}}]{Modi2011}%
  \BibitemOpen
  \bibfield  {author} {\bibinfo {author} {\bibfnamefont {K.}~\bibnamefont
  {Modi}}, \bibinfo {author} {\bibfnamefont {H.}~\bibnamefont {Cable}},
  \bibinfo {author} {\bibfnamefont {M.}~\bibnamefont {Williamson}},\ and\
  \bibinfo {author} {\bibfnamefont {V.}~\bibnamefont {Vedral}},\ }\bibfield
  {title} {\bibinfo {title} {Quantum correlations in mixed-state metrology},\
  }\href {https://doi.org/10.1103/PhysRevX.1.021022} {\bibfield  {journal}
  {\bibinfo  {journal} {Phys. Rev. X}\ }\textbf {\bibinfo {volume} {1}},\
  \bibinfo {pages} {021022} (\bibinfo {year} {2011})}\BibitemShut {NoStop}%
\bibitem [{\citenamefont {Branciard}\ \emph {et~al.}(2012)\citenamefont
  {Branciard}, \citenamefont {Cavalcanti}, \citenamefont {Walborn},
  \citenamefont {Scarani},\ and\ \citenamefont {Wiseman}}]{Branciard2012}%
  \BibitemOpen
  \bibfield  {author} {\bibinfo {author} {\bibfnamefont {C.}~\bibnamefont
  {Branciard}}, \bibinfo {author} {\bibfnamefont {E.~G.}\ \bibnamefont
  {Cavalcanti}}, \bibinfo {author} {\bibfnamefont {S.~P.}\ \bibnamefont
  {Walborn}}, \bibinfo {author} {\bibfnamefont {V.}~\bibnamefont {Scarani}},\
  and\ \bibinfo {author} {\bibfnamefont {H.~M.}\ \bibnamefont {Wiseman}},\
  }\bibfield  {title} {\bibinfo {title} {One-sided device-independent quantum
  key distribution: Security, feasibility, and the connection with steering},\
  }\href {https://doi.org/10.1103/PhysRevA.85.010301} {\bibfield  {journal}
  {\bibinfo  {journal} {Phys. Rev. A}\ }\textbf {\bibinfo {volume} {85}},\
  \bibinfo {pages} {010301} (\bibinfo {year} {2012})}\BibitemShut {NoStop}%
\bibitem [{\citenamefont {He}\ \emph {et~al.}(2015)\citenamefont {He},
  \citenamefont {Rosales-Z\'arate}, \citenamefont {Adesso},\ and\ \citenamefont
  {Reid}}]{He2015}%
  \BibitemOpen
  \bibfield  {author} {\bibinfo {author} {\bibfnamefont {Q.}~\bibnamefont
  {He}}, \bibinfo {author} {\bibfnamefont {L.}~\bibnamefont
  {Rosales-Z\'arate}}, \bibinfo {author} {\bibfnamefont {G.}~\bibnamefont
  {Adesso}},\ and\ \bibinfo {author} {\bibfnamefont {M.~D.}\ \bibnamefont
  {Reid}},\ }\bibfield  {title} {\bibinfo {title} {Secure continuous variable
  teleportation and einstein-podolsky-rosen steering},\ }\href
  {https://doi.org/10.1103/PhysRevLett.115.180502} {\bibfield  {journal}
  {\bibinfo  {journal} {Phys. Rev. Lett.}\ }\textbf {\bibinfo {volume} {115}},\
  \bibinfo {pages} {180502} (\bibinfo {year} {2015})}\BibitemShut {NoStop}%
\bibitem [{\citenamefont {Xiang}\ \emph {et~al.}(2017)\citenamefont {Xiang},
  \citenamefont {Kogias}, \citenamefont {Adesso},\ and\ \citenamefont
  {He}}]{Xiang2017}%
  \BibitemOpen
  \bibfield  {author} {\bibinfo {author} {\bibfnamefont {Y.}~\bibnamefont
  {Xiang}}, \bibinfo {author} {\bibfnamefont {I.}~\bibnamefont {Kogias}},
  \bibinfo {author} {\bibfnamefont {G.}~\bibnamefont {Adesso}},\ and\ \bibinfo
  {author} {\bibfnamefont {Q.}~\bibnamefont {He}},\ }\bibfield  {title}
  {\bibinfo {title} {Multipartite gaussian steering: Monogamy constraints and
  quantum cryptography applications},\ }\href
  {https://doi.org/10.1103/PhysRevA.95.010101} {\bibfield  {journal} {\bibinfo
  {journal} {Phys. Rev. A}\ }\textbf {\bibinfo {volume} {95}},\ \bibinfo
  {pages} {010101} (\bibinfo {year} {2017})}\BibitemShut {NoStop}%
\bibitem [{\citenamefont {Goettems}\ \emph {et~al.}(2021)\citenamefont
  {Goettems}, \citenamefont {Maciel}, \citenamefont {Soares-Pinto},\ and\
  \citenamefont {Duzzioni}}]{Goettems2021}%
  \BibitemOpen
  \bibfield  {author} {\bibinfo {author} {\bibfnamefont {E.~I.}\ \bibnamefont
  {Goettems}}, \bibinfo {author} {\bibfnamefont {T.~O.}\ \bibnamefont
  {Maciel}}, \bibinfo {author} {\bibfnamefont {D.~O.}\ \bibnamefont
  {Soares-Pinto}},\ and\ \bibinfo {author} {\bibfnamefont {E.~I.}\ \bibnamefont
  {Duzzioni}},\ }\bibfield  {title} {\bibinfo {title} {Promoting quantum
  correlations in deterministic quantum computation with a one-qubit model via
  postselection},\ }\href {https://doi.org/10.1103/PhysRevA.103.042416}
  {\bibfield  {journal} {\bibinfo  {journal} {Phys. Rev. A}\ }\textbf {\bibinfo
  {volume} {103}},\ \bibinfo {pages} {042416} (\bibinfo {year}
  {2021})}\BibitemShut {NoStop}%
\bibitem [{\citenamefont {Li}\ and\ \citenamefont {Luo}(2007)}]{Li2007}%
  \BibitemOpen
  \bibfield  {author} {\bibinfo {author} {\bibfnamefont {N.}~\bibnamefont
  {Li}}\ and\ \bibinfo {author} {\bibfnamefont {S.}~\bibnamefont {Luo}},\
  }\bibfield  {title} {\bibinfo {title} {Total versus quantum correlations in
  quantum states},\ }\href {https://doi.org/10.1103/PhysRevA.76.032327}
  {\bibfield  {journal} {\bibinfo  {journal} {Phys. Rev. A}\ }\textbf {\bibinfo
  {volume} {76}},\ \bibinfo {pages} {032327} (\bibinfo {year}
  {2007})}\BibitemShut {NoStop}%
\bibitem [{\citenamefont {Adami}\ and\ \citenamefont {Cerf}(1997)}]{Adami1997}%
  \BibitemOpen
  \bibfield  {author} {\bibinfo {author} {\bibfnamefont {C.}~\bibnamefont
  {Adami}}\ and\ \bibinfo {author} {\bibfnamefont {N.~J.}\ \bibnamefont
  {Cerf}},\ }\bibfield  {title} {\bibinfo {title} {von neumann capacity of
  noisy quantum channels},\ }\href {https://doi.org/10.1103/PhysRevA.56.3470}
  {\bibfield  {journal} {\bibinfo  {journal} {Phys. Rev. A}\ }\textbf {\bibinfo
  {volume} {56}},\ \bibinfo {pages} {3470} (\bibinfo {year}
  {1997})}\BibitemShut {NoStop}%
\bibitem [{\citenamefont {Stratonovich}(1965)}]{Stratonovich1966}%
  \BibitemOpen
  \bibfield  {author} {\bibinfo {author} {\bibfnamefont {R.~L.}\ \bibnamefont
  {Stratonovich}},\ }\bibfield  {title} {\bibinfo {title} {Information capacity
  of a quantum communications channel. i.},\ }\href
  {https://doi.org/10.1007/BF01038470} {\bibfield  {journal} {\bibinfo
  {journal} {I. Soviet Radiophysics}\ }\textbf {\bibinfo {volume} {8}},\
  \bibinfo {pages} {82} (\bibinfo {year} {1965})}\BibitemShut {NoStop}%
\bibitem [{\citenamefont {Meystre}\ and\ \citenamefont
  {Scully}(1983)}]{Zurek1983}%
  \BibitemOpen
  \bibfield  {author} {\bibinfo {author} {\bibfnamefont {P.}~\bibnamefont
  {Meystre}}\ and\ \bibinfo {author} {\bibfnamefont {M.~O.}\ \bibnamefont
  {Scully}},\ }\href@noop {} {\emph {\bibinfo {title} {Quantum optics,
  Experimental gravity, and measurement theory}}}\ (\bibinfo  {publisher}
  {Springer New York},\ \bibinfo {year} {1983})\BibitemShut {NoStop}%
\bibitem [{\citenamefont {Barnett}\ and\ \citenamefont
  {Phoenix}(1989)}]{Barnett1989}%
  \BibitemOpen
  \bibfield  {author} {\bibinfo {author} {\bibfnamefont {S.~M.}\ \bibnamefont
  {Barnett}}\ and\ \bibinfo {author} {\bibfnamefont {S.~J.~D.}\ \bibnamefont
  {Phoenix}},\ }\bibfield  {title} {\bibinfo {title} {Entropy as a measure of
  quantum optical correlation},\ }\href
  {https://doi.org/10.1103/PhysRevA.40.2404} {\bibfield  {journal} {\bibinfo
  {journal} {Phys. Rev. A}\ }\textbf {\bibinfo {volume} {40}},\ \bibinfo
  {pages} {2404} (\bibinfo {year} {1989})}\BibitemShut {NoStop}%
\bibitem [{\citenamefont {Barnett}\ and\ \citenamefont
  {Phoenix}(1991)}]{Barnett1991}%
  \BibitemOpen
  \bibfield  {author} {\bibinfo {author} {\bibfnamefont {S.~M.}\ \bibnamefont
  {Barnett}}\ and\ \bibinfo {author} {\bibfnamefont {S.~J.~D.}\ \bibnamefont
  {Phoenix}},\ }\bibfield  {title} {\bibinfo {title} {Information theory,
  squeezing, and quantum correlations},\ }\href
  {https://doi.org/10.1103/PhysRevA.44.535} {\bibfield  {journal} {\bibinfo
  {journal} {Phys. Rev. A}\ }\textbf {\bibinfo {volume} {44}},\ \bibinfo
  {pages} {535} (\bibinfo {year} {1991})}\BibitemShut {NoStop}%
\bibitem [{\citenamefont {Kumar}(2024)}]{Kumar2024}%
  \BibitemOpen
  \bibfield  {author} {\bibinfo {author} {\bibfnamefont {A.}~\bibnamefont
  {Kumar}},\ }\bibfield  {title} {\bibinfo {title} {Family of quantum mutual
  information and interaction information in multiparty quantum systems},\
  }\bibfield  {journal} {\bibinfo  {journal} {arXiv:2407.16365}\ }\href
  {https://doi.org/10.48550/arXiv.2407.16365} {10.48550/arXiv.2407.16365}
  (\bibinfo {year} {2024})\BibitemShut {NoStop}%
\bibitem [{\citenamefont {Alipour}\ \emph {et~al.}(2020)\citenamefont
  {Alipour}, \citenamefont {Tuohino}, \citenamefont {Rezakhani},\ and\
  \citenamefont {Ala-Nissila}}]{Alipour2020}%
  \BibitemOpen
  \bibfield  {author} {\bibinfo {author} {\bibfnamefont {S.}~\bibnamefont
  {Alipour}}, \bibinfo {author} {\bibfnamefont {S.}~\bibnamefont {Tuohino}},
  \bibinfo {author} {\bibfnamefont {A.~T.}\ \bibnamefont {Rezakhani}},\ and\
  \bibinfo {author} {\bibfnamefont {T.}~\bibnamefont {Ala-Nissila}},\
  }\bibfield  {title} {\bibinfo {title} {Unreliability of mutual information as
  a measure for variations in total correlations},\ }\href
  {https://doi.org/10.1103/PhysRevA.101.042311} {\bibfield  {journal} {\bibinfo
   {journal} {Phys. Rev. A}\ }\textbf {\bibinfo {volume} {101}},\ \bibinfo
  {pages} {042311} (\bibinfo {year} {2020})}\BibitemShut {NoStop}%
\bibitem [{\citenamefont {Tserkis}\ \emph {et~al.}(2023)\citenamefont
  {Tserkis}, \citenamefont {Assad}, \citenamefont {Lam},\ and\ \citenamefont
  {Narang}}]{Tserkis2023}%
  \BibitemOpen
  \bibfield  {author} {\bibinfo {author} {\bibfnamefont {S.}~\bibnamefont
  {Tserkis}}, \bibinfo {author} {\bibfnamefont {S.~M.}\ \bibnamefont {Assad}},
  \bibinfo {author} {\bibfnamefont {P.~K.}\ \bibnamefont {Lam}},\ and\ \bibinfo
  {author} {\bibfnamefont {P.}~\bibnamefont {Narang}},\ }\bibfield  {title}
  {\bibinfo {title} {Quantifying total correlations in quantum systems through
  the pearson correlation coefficient},\ }\bibfield  {journal} {\bibinfo
  {journal} {arXiv:2306.14458}\ }\href
  {https://doi.org/10.48550/arXiv.2306.14458} {10.48550/arXiv.2306.14458}
  (\bibinfo {year} {2023})\BibitemShut {NoStop}%
\bibitem [{\citenamefont {Virmani}\ and\ \citenamefont
  {Plenio}(2000)}]{Virmani2000}%
  \BibitemOpen
  \bibfield  {author} {\bibinfo {author} {\bibfnamefont {S.}~\bibnamefont
  {Virmani}}\ and\ \bibinfo {author} {\bibfnamefont {M.}~\bibnamefont
  {Plenio}},\ }\bibfield  {title} {\bibinfo {title} {Ordering states with
  entanglement measures},\ }\href
  {https://doi.org/10.1016/S0375-9601(00)00157-2} {\bibfield  {journal}
  {\bibinfo  {journal} {Phys. Lett. A}\ }\textbf {\bibinfo {volume} {268}},\
  \bibinfo {pages} {31} (\bibinfo {year} {2000})}\BibitemShut {NoStop}%
\bibitem [{\citenamefont {Henderson}\ and\ \citenamefont
  {Vedral}(2001)}]{Henderson2001}%
  \BibitemOpen
  \bibfield  {author} {\bibinfo {author} {\bibfnamefont {L.}~\bibnamefont
  {Henderson}}\ and\ \bibinfo {author} {\bibfnamefont {V.}~\bibnamefont
  {Vedral}},\ }\bibfield  {title} {\bibinfo {title} {Classical, quantum and
  total correlations},\ }\href {https://doi.org/10.1088/0305-4470/34/35/315}
  {\bibfield  {journal} {\bibinfo  {journal} {J. Phys. A Math.}\ }\textbf
  {\bibinfo {volume} {34}},\ \bibinfo {pages} {6899} (\bibinfo {year}
  {2001})}\BibitemShut {NoStop}%
\bibitem [{\citenamefont {Zhou}\ \emph {et~al.}(2006)\citenamefont {Zhou},
  \citenamefont {Zeng}, \citenamefont {Xu},\ and\ \citenamefont
  {You}}]{Zhou2006}%
  \BibitemOpen
  \bibfield  {author} {\bibinfo {author} {\bibfnamefont {D.~L.}\ \bibnamefont
  {Zhou}}, \bibinfo {author} {\bibfnamefont {B.}~\bibnamefont {Zeng}}, \bibinfo
  {author} {\bibfnamefont {Z.}~\bibnamefont {Xu}},\ and\ \bibinfo {author}
  {\bibfnamefont {L.}~\bibnamefont {You}},\ }\bibfield  {title} {\bibinfo
  {title} {Multiparty correlation measure based on the cumulant},\ }\href
  {https://doi.org/10.1103/PhysRevA.74.052110} {\bibfield  {journal} {\bibinfo
  {journal} {Phys. Rev. A}\ }\textbf {\bibinfo {volume} {74}},\ \bibinfo
  {pages} {052110} (\bibinfo {year} {2006})}\BibitemShut {NoStop}%
\bibitem [{\citenamefont {Vedral}(2003)}]{Vedral2003}%
  \BibitemOpen
  \bibfield  {author} {\bibinfo {author} {\bibfnamefont {V.}~\bibnamefont
  {Vedral}},\ }\bibfield  {title} {\bibinfo {title} {Classical correlations and
  entanglement in quantum measurements},\ }\href
  {https://doi.org/10.1103/PhysRevLett.90.050401} {\bibfield  {journal}
  {\bibinfo  {journal} {Phys. Rev. Lett.}\ }\textbf {\bibinfo {volume} {90}},\
  \bibinfo {pages} {050401} (\bibinfo {year} {2003})}\BibitemShut {NoStop}%
\bibitem [{\citenamefont {Peres}(1996)}]{Peres1996}%
  \BibitemOpen
  \bibfield  {author} {\bibinfo {author} {\bibfnamefont {A.}~\bibnamefont
  {Peres}},\ }\bibfield  {title} {\bibinfo {title} {Separability criterion for
  density matrices},\ }\href {https://doi.org/10.1103/PhysRevLett.77.1413}
  {\bibfield  {journal} {\bibinfo  {journal} {Phys. Rev. Lett.}\ }\textbf
  {\bibinfo {volume} {77}},\ \bibinfo {pages} {1413} (\bibinfo {year}
  {1996})}\BibitemShut {NoStop}%
\bibitem [{\citenamefont {Horodecki}\ \emph {et~al.}(1996)\citenamefont
  {Horodecki}, \citenamefont {Horodecki},\ and\ \citenamefont
  {Horodecki}}]{Horodecki1996}%
  \BibitemOpen
  \bibfield  {author} {\bibinfo {author} {\bibfnamefont {R.}~\bibnamefont
  {Horodecki}}, \bibinfo {author} {\bibfnamefont {M.}~\bibnamefont
  {Horodecki}},\ and\ \bibinfo {author} {\bibfnamefont {P.}~\bibnamefont
  {Horodecki}},\ }\bibfield  {title} {\bibinfo {title} {Teleportation, bell's
  inequalities and inseparability},\ }\href
  {https://doi.org/10.1016/0375-9601(96)00639-1} {\bibfield  {journal}
  {\bibinfo  {journal} {Phys. Lett. A}\ }\textbf {\bibinfo {volume} {222}},\
  \bibinfo {pages} {21} (\bibinfo {year} {1996})}\BibitemShut {NoStop}%
\bibitem [{\citenamefont {Schumacher}\ and\ \citenamefont
  {Alber}(2023)}]{Schumacher2023}%
  \BibitemOpen
  \bibfield  {author} {\bibinfo {author} {\bibfnamefont {M.}~\bibnamefont
  {Schumacher}}\ and\ \bibinfo {author} {\bibfnamefont {G.}~\bibnamefont
  {Alber}},\ }\bibfield  {title} {\bibinfo {title} {Detection of typical
  bipartite entanglement by local generalized measurements},\ }\href
  {https://doi.org/10.1103/PhysRevA.108.042424} {\bibfield  {journal} {\bibinfo
   {journal} {Phys. Rev. A}\ }\textbf {\bibinfo {volume} {108}},\ \bibinfo
  {pages} {042424} (\bibinfo {year} {2023})}\BibitemShut {NoStop}%
\bibitem [{\citenamefont {Umegaki}(1962)}]{Umegaki1962}%
  \BibitemOpen
  \bibfield  {author} {\bibinfo {author} {\bibfnamefont {H.}~\bibnamefont
  {Umegaki}},\ }\bibfield  {title} {\bibinfo {title} {{Conditional expectation
  in an operator algebra. IV. Entropy and information}},\ }\href
  {https://doi.org/10.2996/kmj/1138844604} {\bibfield  {journal} {\bibinfo
  {journal} {Kodai Math. Sem. Rep.}\ }\textbf {\bibinfo {volume} {14}},\
  \bibinfo {pages} {59 } (\bibinfo {year} {1962})}\BibitemShut {NoStop}%
\bibitem [{\citenamefont {Cover}\ and\ \citenamefont
  {Thomas}(2006)}]{Cover2006}%
  \BibitemOpen
  \bibfield  {author} {\bibinfo {author} {\bibfnamefont {T.~M.}\ \bibnamefont
  {Cover}}\ and\ \bibinfo {author} {\bibfnamefont {J.~A.}\ \bibnamefont
  {Thomas}},\ }\href@noop {} {\emph {\bibinfo {title} {Elements of Information
  Theory}}}\ (\bibinfo  {publisher} {Wiley, New York},\ \bibinfo {year}
  {2006})\BibitemShut {NoStop}%
\bibitem [{\citenamefont {Orthey}\ and\ \citenamefont
  {Angelo}(2022)}]{Orthey2022}%
  \BibitemOpen
  \bibfield  {author} {\bibinfo {author} {\bibfnamefont {A.~C.}\ \bibnamefont
  {Orthey}}\ and\ \bibinfo {author} {\bibfnamefont {R.~M.}\ \bibnamefont
  {Angelo}},\ }\bibfield  {title} {\bibinfo {title} {Quantum realism:
  Axiomatization and quantification},\ }\href
  {https://doi.org/10.1103/PhysRevA.105.052218} {\bibfield  {journal} {\bibinfo
   {journal} {Phys. Rev. A}\ }\textbf {\bibinfo {volume} {105}},\ \bibinfo
  {pages} {052218} (\bibinfo {year} {2022})}\BibitemShut {NoStop}%
\bibitem [{\citenamefont {Petz}(1986)}]{Petz1986}%
  \BibitemOpen
  \bibfield  {author} {\bibinfo {author} {\bibfnamefont {D.}~\bibnamefont
  {Petz}},\ }\bibfield  {title} {\bibinfo {title} {Quasi-entropies for finite
  quantum systems},\ }\href {https://doi.org/10.1016/0034-4877(86)90067-4}
  {\bibfield  {journal} {\bibinfo  {journal} {Rep. Math. Phys.}\ }\textbf
  {\bibinfo {volume} {23}},\ \bibinfo {pages} {57} (\bibinfo {year}
  {1986})}\BibitemShut {NoStop}%
\bibitem [{\citenamefont {Mosonyi}\ and\ \citenamefont
  {Hiai}(2011)}]{Mosonyi2011}%
  \BibitemOpen
  \bibfield  {author} {\bibinfo {author} {\bibfnamefont {M.}~\bibnamefont
  {Mosonyi}}\ and\ \bibinfo {author} {\bibfnamefont {F.}~\bibnamefont {Hiai}},\
  }\bibfield  {title} {\bibinfo {title} {On the quantum rényi relative
  entropies and related capacity formulas},\ }\href
  {https://doi.org/10.1109/TIT.2011.2110050} {\bibfield  {journal} {\bibinfo
  {journal} {IEEE Transactions on Information Theory}\ }\textbf {\bibinfo
  {volume} {57}},\ \bibinfo {pages} {2474} (\bibinfo {year}
  {2011})}\BibitemShut {NoStop}%
\bibitem [{\citenamefont {Abe}(2003)}]{Abe2003}%
  \BibitemOpen
  \bibfield  {author} {\bibinfo {author} {\bibfnamefont {S.}~\bibnamefont
  {Abe}},\ }\bibfield  {title} {\bibinfo {title} {Monotonic decrease of the
  quantum nonadditive divergence by projective measurements},\ }\href
  {https://doi.org/10.1016/S0375-9601(03)00682-0} {\bibfield  {journal}
  {\bibinfo  {journal} {Phys. Lett. A}\ }\textbf {\bibinfo {volume} {312}},\
  \bibinfo {pages} {336} (\bibinfo {year} {2003})}\BibitemShut {NoStop}%
\bibitem [{\citenamefont {Rastegin}(2016)}]{Rastegin2016}%
  \BibitemOpen
  \bibfield  {author} {\bibinfo {author} {\bibfnamefont {A.~E.}\ \bibnamefont
  {Rastegin}},\ }\bibfield  {title} {\bibinfo {title} {Quantum-coherence
  quantifiers based on the tsallis relative $\ensuremath{\alpha}$ entropies},\
  }\href {https://doi.org/10.1103/PhysRevA.93.032136} {\bibfield  {journal}
  {\bibinfo  {journal} {Phys. Rev. A}\ }\textbf {\bibinfo {volume} {93}},\
  \bibinfo {pages} {032136} (\bibinfo {year} {2016})}\BibitemShut {NoStop}%
\bibitem [{\citenamefont {Hiai}\ \emph {et~al.}(2011)\citenamefont {Hiai},
  \citenamefont {Mosonyi}, \citenamefont {Petz},\ and\ \citenamefont
  {B{\'e}ny}}]{Hiai2011}%
  \BibitemOpen
  \bibfield  {author} {\bibinfo {author} {\bibfnamefont {F.}~\bibnamefont
  {Hiai}}, \bibinfo {author} {\bibfnamefont {M.}~\bibnamefont {Mosonyi}},
  \bibinfo {author} {\bibfnamefont {D.}~\bibnamefont {Petz}},\ and\ \bibinfo
  {author} {\bibfnamefont {C.}~\bibnamefont {B{\'e}ny}},\ }\bibfield  {title}
  {\bibinfo {title} {Quantum f-divergences and error correction},\ }\href
  {https://doi.org/10.1142/S0129055X11004412} {\bibfield  {journal} {\bibinfo
  {journal} {Rev. Math. Phys.}\ }\textbf {\bibinfo {volume} {23}},\ \bibinfo
  {pages} {691} (\bibinfo {year} {2011})}\BibitemShut {NoStop}%
\bibitem [{\citenamefont {Bhatia}(2013)}]{Bhatia2013}%
  \BibitemOpen
  \bibfield  {author} {\bibinfo {author} {\bibfnamefont {R.}~\bibnamefont
  {Bhatia}},\ }\href@noop {} {\emph {\bibinfo {title} {\textnormal{Matrix
  analysis}}}}\ (\bibinfo  {publisher} {Springer Science \& Business Media},\
  \bibinfo {year} {2013})\BibitemShut {NoStop}%
\bibitem [{\citenamefont {Pearson}(1895)}]{Pearson1895}%
  \BibitemOpen
  \bibfield  {author} {\bibinfo {author} {\bibfnamefont {K.}~\bibnamefont
  {Pearson}},\ }\bibfield  {title} {\bibinfo {title} {Note on regression and
  inheritance in the case of two parents},\ }\href
  {https://doi.org/10.1098/rspl.1895.0041} {\bibfield  {journal} {\bibinfo
  {journal} {R. Soc. Lond.}\ }\textbf {\bibinfo {volume} {58}},\ \bibinfo
  {pages} {240} (\bibinfo {year} {1895})}\BibitemShut {NoStop}%
\bibitem [{\citenamefont {R{\'e}nyi}\ and\ \citenamefont
  {Vekerdi}(1970)}]{Renyi1970}%
  \BibitemOpen
  \bibfield  {author} {\bibinfo {author} {\bibfnamefont {A.}~\bibnamefont
  {R{\'e}nyi}}\ and\ \bibinfo {author} {\bibfnamefont {L.}~\bibnamefont
  {Vekerdi}},\ }\href@noop {} {\emph {\bibinfo {title} {Probability Theory}}}\
  (\bibinfo  {publisher} {North-Holland Publishing Company},\ \bibinfo {year}
  {1970})\BibitemShut {NoStop}%
\bibitem [{\citenamefont {Beigi}(2013)}]{Beigi2013}%
  \BibitemOpen
  \bibfield  {author} {\bibinfo {author} {\bibfnamefont {S.}~\bibnamefont
  {Beigi}},\ }\bibfield  {title} {\bibinfo {title} {A new quantum data
  processing inequality},\ }\href {https://doi.org/10.1063/1.4818985}
  {\bibfield  {journal} {\bibinfo  {journal} {J. Math. Phys.}\ }\textbf
  {\bibinfo {volume} {54}},\ \bibinfo {pages} {082202} (\bibinfo {year}
  {2013})}\BibitemShut {NoStop}%
\bibitem [{\citenamefont {Beigi}\ and\ \citenamefont
  {Rahimi-Keshari}(2023)}]{Beigi2023}%
  \BibitemOpen
  \bibfield  {author} {\bibinfo {author} {\bibfnamefont {S.}~\bibnamefont
  {Beigi}}\ and\ \bibinfo {author} {\bibfnamefont {S.}~\bibnamefont
  {Rahimi-Keshari}},\ }\bibfield  {title} {\bibinfo {title} {Quantum maximal
  correlation for gaussian states},\ }\href
  {https://doi.org/10.1103/PhysRevA.108.062419} {\bibfield  {journal} {\bibinfo
   {journal} {Phys. Rev. A}\ }\textbf {\bibinfo {volume} {108}},\ \bibinfo
  {pages} {062419} (\bibinfo {year} {2023})}\BibitemShut {NoStop}%
\bibitem [{\citenamefont {Gour}(2021)}]{Gour2021}%
  \BibitemOpen
  \bibfield  {author} {\bibinfo {author} {\bibfnamefont {G.}~\bibnamefont
  {Gour}},\ }\bibfield  {title} {\bibinfo {title} {Uniqueness and optimality of
  dynamical extensions of divergences},\ }\href
  {https://doi.org/10.1103/PRXQuantum.2.010313} {\bibfield  {journal} {\bibinfo
   {journal} {PRX Quantum}\ }\textbf {\bibinfo {volume} {2}},\ \bibinfo {pages}
  {010313} (\bibinfo {year} {2021})}\BibitemShut {NoStop}%
\bibitem [{\citenamefont {Maziero}\ \emph {et~al.}(2011)\citenamefont
  {Maziero}, \citenamefont {Celeri},\ and\ \citenamefont
  {Serra}}]{Maziero2011}%
  \BibitemOpen
  \bibfield  {author} {\bibinfo {author} {\bibfnamefont {J.}~\bibnamefont
  {Maziero}}, \bibinfo {author} {\bibfnamefont {L.~C.}\ \bibnamefont
  {Celeri}},\ and\ \bibinfo {author} {\bibfnamefont {R.~M.}\ \bibnamefont
  {Serra}},\ }\bibfield  {title} {\bibinfo {title} {Symmetry aspects of quantum
  discord},\ }\bibfield  {journal} {\bibinfo  {journal} {arXiv:1004.2082}\
  }\href {https://doi.org/10.48550/arXiv.1004.2082} {10.48550/arXiv.1004.2082}
  (\bibinfo {year} {2011})\BibitemShut {NoStop}%
\end{thebibliography}%

\end{document}